\documentstyle[12pt]{article}

\author{Hao Chen\\
Department of Mathematics\\
Zhongshan University\\
Guangzhou,Guangdong 510275\\
People's Republic of China}
\title{Quantum entanglement without eigenvalue spectra: multipartite case}
\date{July,2001}

\begin{document}

\maketitle
\begin{abstract}
We introduce algebraic sets in the products of the complex projective spaces for the mixed states in multipartite quantum systems, which are independent of their eigenvalues and only measure the ``position'' of their eigenvectors,  as their non-local invariants (ie., remaining invariant after local unitary transformation). These invariants are naturally arised from the physical consideration of checking multipartite mixed states by measuring them with the multipartite separable pure states. The algebraic sets have to be the sum of linear subspaces if the multipartite  mixed state is separable, and thus we give a new criterion of separability. A continuous family of 4-party mixed states, whose members are separable for any $2:2$ cut and entangled for any $1:3$ cut (thus bound entanglement if 4 parties are isolated), is constructed and studied by our invariants and separability criterion. Examples of LOCC-incomparable entangled tripartite pure states are also given to show that it is hopeless to characterize the entanglement proporties of multipartite pure states by the eigenvalue spectra of their partial traces. We also prove that at least $n^2+n-1$ terms of separable pure states, which are orthogonal in some sence, are needed to write a generic pure state in $H_A^{n^2} \otimes H_B^{n^2} \otimes H_C^{n^2}$ as a linear combination of them.

\end{abstract}

Entanglement, first noted by Einstein, Podolsky and Rosen [1] and Schrodinger [2], is an essential feature of quantum mechanics. Recently it has been realized that entanglement is a useful resource for various kinds of quantum information processing, including quantum state teleportation ([3]), cryptographic key distribution ([3],[4]), quantum computation ([5]), etc., see [6] and [7].\\

Entanglement of bipartite quantum systems, ie., entanglement of pure and mixed states in $H=H_A^m \otimes H_B^n$, has been a matter of intensive research, see [7] for a survey. It has been realized that the entanglement of pure tripartite quantum states is not a trivial extension of the entanglement of bipartite systems ([8],[9]). Recently Bennett etc., [10] studied the exact and asymptotic entanglement measure of multipartite pure states, which showed  essential difference to that of  bipartite pure states. On the other hand Acin, etc., [11] proved a generalization of Schmidt decomposition for pure triqubit states, which seems impossible to be generalized to arbitrary multipartite case (see Theorem 5 in this paper). Basically, the understanding of multipartite quantum entanglement for both pure and mixed states, is much less advanced.\\

From the point view of quantum entanglement, states in multipartite quantum systems $H=H_{A_1}^{m_1} \otimes \cdots \otimes H_{A_n}^{m_n}$ are completely equivalent if they can be transformed by local unitary transfomations (ie., $U_{A_1} \otimes \cdots \otimes U_{A_n}$, where $U_{A_i}$ is unitary transformation of $H_{A_i}$). The property of states being separable or entangled is clearly preserved under local unitary transformations. Any good measure of multipartite entanglement must be invariant under local unitary transformations ([12]). It is obvious that eigenvalues of $tr_{A_{i_1}...A_{i_k}}(\rho)$, for any indices $i_1,...,i_k \in \{1,...,n\}$ and any state $\rho$ in $H$, are invariant under local unitary transformations.  Many known invariants [12] and entanglement monotone (see [13]) are more or less related to eigenvalue spectra of states or their partial traces.\\

It is clear that any separability criterion for bipartite mixed states , such as Peres PPT criterion [14] and Horodecki range criterion [7], can be applied to multipartite mixed states for their separability under various cuts. For example, from Peres PPT criterion, a separable multipartite mixed state necessarily have all its partial transpositions positive. In [15], Horodeckis proved a separability criterion for multipartite mixed states by the using of linear maps. Based on range criterion, a systematic way to construct both bipartite and multipartite PPT mixed states (thus bound entanglement) from unextendible product bases was eastablished in [16]. Classification of triqubit mixed states inspired by Acin, etc., [11] was studied in [17].\\ 

In our previous work [18] we introduced  algebraic sets in complex projective spaces for states in bipartite quantum systems, which are independent of the eigenvalues of the states and only measure the ``position'' of eigenvectors, as their non-local invariants. These invariants are naturally from the physical consideration to measure the bipartite mixed states by separable pure states. An ``eigenvalue-free'' separability criterion is also proved, which asserts that the algebraic sets have to be the sum of linear subspaces if the state is separable. This revealed that a quite large part of quantum entanglement is independent of eigenvalue spectra and only dependes on eigenvectors.\\

The algebraic set invariants and the separability criterion based on these invariants in [18] can be extended naturally to multipartite mixed states. For any multipartite mixed state, we measure it by multipartite separable pure states and consider the ``degenerating locus''. From this motivation we introduce algebraic sets (ie., zero locus of several homogeneous multi-variable polynomials, see [19]) in the products of complex projective spaces for the multipartite mixed states, which are independent of the eigenvalues of the mixed states and only measure the ``geometric position'' of eigenvectors. These algebraic sets are invariants of the mixed states under local unitary transformations, and thus many numerical algebraic-geometric invariants (such as dimensions, number of irreducible components) and Hermitian differential geometric invariants (with the product Fubini-Study metric on the products of complex projective spaces,  such as volumes, curvatures) of these algebraic sets are automatically invariant under local unitary transformations. In this way many candidates for good entanglement measure or potentially entanglement monotone independent of eigenvalues of mixed states are offerd. Another important aspect is that these algebraic sets can be easily calculated. Based on these algebraic sets we prove a new separability criterion (independent of eigenvalues) which asserts that the algebraic sets have to be the sum of the products of linear subspaces if the multipartite mixed state is separable. For any entangled mixed state violating our criterion, the mixed states with the same eigenvectors and arbitrary eigenvalues are also entangled, ie., our criterion always detects continuous family of entangled multipartite mixed states. \\

Based on our new separabilty criterion, a continuous family of 4 quibit mixed state is constructed as a generalizationof Smolin's mixed state in [20], each mixed state in this family is separable under any $2:2$ cut and entangled under any $1:3$ cut, thus they are bound entanglement if 4 parties are isolated (Example 1). Since our invariants can be computed easily and can be used to distinguish inequivalent mixed states under local unitary transformations, it is proved that the ``generic'' members of this continuous family of mixed states are inequivalent under local unitary transformations, thus these 4 qubit mixed states are continuous many distinct bound entangled mixed states. Examples of LOCC-incomparable enatngled tripartite pure states are also constructed to show it is hopeless to characterize the entanglement properties of multipartite pure states by the eigenvalue vectors of their partial traces, actually the eigenvalue vectors of partial traces of the tripartite pure states in example 2 are constant. At last we proved that at least $n^2+n-1$ pure separable states, which are orthogonal in some sence, are needed to write any generic pure state in $H_A^{n^2} \otimes H_B^{n^2} \otimes H_C^{n^2}$ as a linear combination of them. This is a remarkable difference to Schmidt decompositions of bipartite pure states.\\

For the algebraic geometry used in this paper, we refer to the nice book [19].\\

We introduce the algebraic sets of the mixed states and prove the results for tripartite case.  Then the multipartite case is similar and we just generalize directly.\\

Let $H=H_{A}^m \otimes H_{B}^n \otimes H_{C}^l$ and the standard orthogonal base is $|ijk>$, where, $i=1,...,m$,$j=1,...,n$ and $k=1,...,l$, and $\rho$ is a mixed state on $H$. We represent the matrix of $\rho$ in the base $\{|111>,...|11l>,...,|mn1>,...,|mnl>\}$ as
 $\rho=(\rho_{ij,i'j'})_{1 \leq i ,i' \leq m, 1 \leq j ,j' \leq n}$, and  
$\rho_{ij,i'j'}$ is a $l \times l$ matrix. Consider $H$ as a bipartite system as $H=(H_{A}^m \otimes H_{B}^n) \otimes H_{C}^l$, then we have $V_{AB}^k(\rho)=\{(r_{11},...,r_{mn}) \in C^{mn}:rank( \Sigma r_{ij}r_{i'j'}^{*} \rho_{ij,i'j'})\leq k \}$ defined as in [18]. This set is actually the ``degenerating locus'' of the Hermitian bilinear form $<\phi_{12}|\rho|\phi_{12}>$ on $H_C^l$ for the given pure state $\phi_{12} =\Sigma_{i,j}^{m,n} r_{ij} |ij> \in P(H_A^m \otimes H_B^n)$. When the finer cut A:B:C is considered, it is natural to take $\phi_{12}$ as a separable pure state $\phi_{12}=\phi_1 \otimes \phi_2$, ie., there exist $\phi_1=\Sigma_i r_i^1 |i> \in P(H_A^m),\phi_2=\Sigma_j r_j^2 |j> \in P(H_B^n)$ such that $r_{ij}=r_i^1r_j^2$. In this way the tripartite mixed state $\rho$ is measured by tripartite separable pure states $\phi_1 \otimes \phi_2 \otimes  \phi_3$. Thus it is natural we define $V_{A:B}^k(\rho)$ as follows. It is the ``degenerating locus'' of the bilinear form $<\phi_1 \otimes \phi_2 |\rho|\phi_1 \otimes \phi_2>$ on $H_C^l$.\\

{\bf Definition 1.}{\em  Let $\phi:CP^{m-1} \times CP^{n-1} \rightarrow CP^{mn-1}$  be the mapping defined by\\

$$
\begin{array}{ccccccccc}
\phi(r_1^1,...r_m^1,r_1^2,...,r_n^2)=(r_1^1r_1^2,...,r_i^1r_j^2,...r_m^1r_n^2)
\end{array}
(1)
$$

(ie., $r_{ij}=r_i^1 r_j^2$ is introduced.)

Then $V_{A:B}^k (\rho)$ is defined as the preimage $\phi^{-1}(V_{AB}^k(\rho))$.}\\

Similarly $V_{B:C}^k(\rho),V_{A:C}^k(\rho)$ can be defined. In the following statement we just state the result for $V_{A:B}^k(\rho)$. The conclusion holds similarly for other $V's$.\\

From this definition and Theorem 2 in [18] we immediately have the following result.\\

{\bf Theorem 1.} {\em $V_{A:B}^k(\rho)$ is an algebraic set in $CP^{m-1} \times CP^{n-1}$.}\\

{\bf Theorem 2.}{\em  Let $T=U_{A} \otimes U_{B} \otimes U_{C}$, where $U_{A},U_{B}$ and $U_{C}$ are unitary transformations of $H_A^m, H_B^n, H_C^l$,be a local unitary transformations of $H$. Then $V_{A:B}^k(T(\rho))=U_{A}^{-1} \times U_{B}^{-1}(V_{A:B}^k(\rho))$.}\\

{\bf Proof.} Let $U_{A}=(u_{ij}^{A})_{1 \leq i \leq m,1 \leq j \leq m}$, $U_{B}=(u_{ij}^{B})_{1 \leq i \leq n, 1 \leq j \leq n}$ and $U_{C}=(u_{ij}^{C})_{1 \leq i \leq l, 1 \leq j \leq l}$, be the matrix in the standard orthogonal bases.\\

Recall the proof of Theorem 1 in [18], we have $V_{AB}^k(T(\rho))=(U_{A} \times U_{B})^{-1}(V_{AB}^k(\rho))$ under the coordinate change\\

$$
\begin{array}{cccccccc} 
r_{kw}'=\Sigma_{ij} r_{ij} u_{ik}^{A} u_{jw}^{B}\\
=\Sigma_{ij}r_i^1 r_j^2 u_{ik}^{A} u_{jw}^{B}\\
=\Sigma_{ij} (r_{i}^1 u_{ik}^A) (r_{j}^2 u_{jw}^B)\\
=(\Sigma_i r_i^1 u_{ik}^A)(\Sigma_j r_j^2 u_{jw}^B)
\end{array}
(2)
$$

for $k=1,...,m,w=1,...,n$. Thus our conclusion follows from the definition.\\

 Since $U_{A}^{-1} \times U_{B}^{-1}$ certainly preserves the  (product) Fubini-Study metric of $CP^{m-1} \times CP^{n-1}$, we know that all metric properties of $V_{A:B}^k(\rho)$ are preserved when the local unitary transformations are applied to the mixed state $\rho$.\\

In the following statement we give a separability criterion of the mixed state $\rho$ under the cut A:B:C. The term ``a linear subspace of $CP^{m-1} \times CP^{n-1}$'' means the product of a linear subspace in $CP^{m-1}$ and a linear subspace in $CP^{n-1}$.\\

{\bf Theorem 3.}{\em If $\rho$ is a separable mixed state on $H=H_{A}^m \otimes H_{B}^{n} \otimes H_{C}^l$ under the cut A:B:C, $V_{A:B}^k(\rho)$ is a linear subset of $CP^{m-1} \times CP^{n-1}$, ie., it is the sum of the linear subspaces.}\\

{\bf Proof.} We first consider the separability of $\rho$ under the cut AB:C,ie., $\rho= \Sigma_{f=1}^g p_f P_{a_f \otimes b_f}$, where $a_f \in H_{A}^m \otimes H_{B}^n$ and $b_f \in H_{C}^l$ for $f=1,...,g$. Consider the separability of $\rho$ under the cut A:B:C, we have  $a_f=a_f' \otimes a_f''$ , $a_f' \in H_{A}^m, a_f'' \in H_{B}^n$. Let $a_f=(a_f^1,...,a_f^{mn}), a_f'=(a_f'^1,...,a_f'^m)$ and $a_f''(a_f''^1,...,a_f''^n)$ be the coordinate forms with the standard orthogonal basis $\{|ij>\}$, $\{|i>\}$ and $\{|j>\}$ respectively, we have that $a_f^{ij}=a_f'^i a_f''^j$. Recall the proof of Theorem 3 in [18], the diagonal entries of $G$ in the proof of Theorem 3 in [18] are\\

$$
\begin{array}{cccccccc}
\Sigma_{ij}r_{ij} a_f^{ij}=\\
\Sigma_{ij} r_i^1 a_f'^i r_j^2 a_f''^j=\\
(\Sigma_i r_i^1 a_f'^i)(\Sigma_j r_j^1 a_f''^j)
\end{array}
(3)
$$

Thus as argued in the proof of Theorem 3 of [18], $V_{A:B}^k(\rho)$ has to be the zero locus of the multiplications of the linear forms in (3). The conclusion is proved.\\

For the mixed state $\rho$ in the multipartite system $H=H_{A_1}^{m_1} \otimes \cdots \otimes H_{A_k}^{m_k}$, we want to study the entanglement under the cut $ A_{i_1}:A_{i_2}:...:A_{i_l}:(A_{j_1}...A_{j_{k-l}})$, where $\{i_1,...,i_l\} \cup \{j_1,...j_{k-l}\}=\{1,...k\}$. We can define the set $V_{A_{i_1}:...:A_{i_l}}^k(\rho)$ similarly. We have the following results.\\

{\bf Theorem 1'.} {\em $V_{A_{i_1}:...:A_{i_l}}^k(\rho)$ is an algebraic set in in $CP^{m_{i_1}-1} \times CP^{m_{i_l}-1}$.}\\

{\bf Theorem 2'.}{\em  Let $T=U_{A_{i_1}} \otimes \cdots \otimes U_{A_{i_l}} \otimes U_{j_1...j_{k-l}}$, where $U_{A_{i_1}},...,U_{A_{i_l}}, U_{j_1...j_{k-l}}$ are unitary transformations of $H_{A_{i_1}},...,H_{A_{i_l}}$, be a local unitary transformations of $H$. Then $V_{A_{i_1}:...:A_{i_l}}^k(T(\rho))=U_{A_{i_1}}^{-1} \times \cdots \times U_{A_{i_l}}^{-1}(V_{A_{i_1}:...:A_{i_l}}^k(\rho))$.}\\

{\bf Theorem 3'.}{\em If $\rho$ is a separable mixed state on $H=H_{A_1}^{m_1} \otimes \cdots \otimes H_{A_k}^{m_k}$ under the cut $ A_{i_1}:A_{i_2}:...:A_{i_l}:(A_{j_1}...A_{j_{k-l}})$, $V_{A_{i_1}:...:A_{i_l}}^k(\rho)$ is a linear subset of $CP^{m_{i_1}-1}  \times ... \times CP^{m_{i_l}-1}$,ie., it is the sum of the linear subspaces.}\\

We now give and study some examples of mixed states based on our above results.\\

The following example, which is a continuous family (depending on 4 parameters) of  mixed state in the four-party quantum system $H_A^2 \otimes H_B^2 \otimes H_C^2 \otimes H_D^2$ and separable for any $2:2$ cut but entangled for any $1:3$ cut,  can be thought as a generalization of Smolin's mixed state in [20].\\

{\bf Example 1.} Let $H=H_{A}^2 \otimes H_{B}^2 \otimes H_{C}^2 \otimes H_{D}^2$ and $h_1,h_2,h_3,h_4$ (understood as row vectors)are 4 mutually orthogonal unit vectors in $C^4$. Consider the $16 \times 4$ matrix $T$ with 16 rows as\\
 $T=(a_1h_1^{\tau},0,0,a_2 h_2^{\tau},0, a_3 h_3^{\tau},a_4 h_4^{\tau},0,0, a_5 h_3^{\tau}, a_6 h_4^{\tau},0, a_7 h_1^{\tau},0,0,a_8 h_2^{\tau})^{\tau}$. Let \\ $\phi'_1,\phi'_2,\phi'_3,\phi'_4$ be 4 vectors in $H$ whose expansions with the base $|0000>,|0001>,|0010>,|0011>,|0100>,|0101>,|0110>,|0111>,|1000>$,\\$|1001>,|1010>,|1011>,|1100>,|1101>,|1110>,|1111> $ are exactly the 4 columns of the matrix $T$ and $\phi_1,\phi_2,\phi_3,\phi_4$ are the normalized unit vectors of $\phi'_1,\phi'_2, \phi'_3, \phi'_4$. Let $\rho=\frac{1}{4}(P_{\phi_1} +P_{\phi_2} +P_{\phi_3} +P_{\phi_4})$.\\

It is easy to check that when $h_1=(1,1,0,0),h_2=(1,-1,0,0), h_3=(0,0,1,1), h_4=(0,0,1,-1)$ and $a_1=a_2=a_3=a_4=1$. It is just the Smolin's mixed state in [20].\\

Now we prove that $\rho$ is invariant under the partial transposes of the cuts AB:CD,AC:BD,AD:BC.\\

Let  the ``representation'' matrix $T=(b_{ijkl})_{i=0,1,j=0,1,k=0,1,l=0,1}$ is the matrix with columns corresponding the expansions of $\phi_1,\phi_2,\phi_3,\phi_4$.Then we can consider that $T=(T_1,T_2,T_3,T_4)^{\tau}$ is blocked matrix of size $4 \times 1$ with each block $T_{ij}=(b_{kl })_{k=0,1,l=0,1}$ a $4 \times 4$ matrix,where $ij=00,01,10,11$. Because $h_1,h_2,h_3,h_4$ are mutually orthogonal unit vectors we can easily check that $T_{ij} (T_{i'j'}^{*})^{\tau}=T_{i'j'} (T_{ij}^{*})^{\tau}$ Thus it is invariant when the partial transpose of the cut AB:CD is applied.\\

With the same methods we can check that $\rho$ is invariant when the partial transposes of the cuts AC:BD, AD:BC  are applied. Hence $\rho$ is PPT under the cuts AB:CD, AC:BD,AD:BC. Thus from a result in [21] we know $\rho$ is separable under these cuts AB:CD, AC:BD,AD:BC.\\

Now we want to prove $\rho$ is entangled under the cut A:BCD by computing $V_{BCD}^1(\rho)$. From the arguments in [18] and this paper, we can check that $V_{BCD}^1(\rho)$ is the locus of the condition: $a_1 h_1 r_{000} + a_2 h_2 r_{011} +a_3 h_3 r_{101} +a_4 h_4 r_{110}$ and $a_7 h_1 r_{100} + a_8 h_2 r_{111} +a_5 h_3 r_{001} +a_6 h_4 r_{010}$ are linear dependent. This is equivalent to the condition that the matrix (6) is of  rank 1.\\

$$
\left(
\begin{array}{cccccc}
a_7 r_{100} & a_8 r_{111} & a_5 r_{001} & a_6 r_{010}\\
a_1 r_{000} & a_2 r_{011} & a_3 r_{101} & a_4 r_{110}
\end{array}
\right)
(6)
$$

From [19] pp. 25-26 we can check that $V_{BCD}^1(\rho)$ is exactly the famous Segre variety in algebraic geometry. It is irreducible and thus cannot be linear. From Theorem 3 in [18], $\rho$ is entangled under the cut A:BCD. Similarly we can prove that $\rho$ is entangled under the cuts B:ACD, C:ABD, D:ABC.\\

Now we compute $V_{A:B}^3(\rho)$. From the arguments in [18] and Definition 1 , it is just the locus of the condition that the vectors $h_1(a_1 r_0^1 r_0^2 +a_7 r_1^1 r_1^2)$, $h_3 (a_3 r_0^1 r_1^2 +a_5 r_1^1 r_0^2)$, $h_4(a_4 r_0^1 r_1^2 +a_6  r_1^1 r_0^2)$, $h_2 (a_2 r_0^1 r_0^2 +a_8 r_1^1 r_1^2)$ are linear dependent. Since $h_1,h_2,h_3,h_4$ are mutually orthogonal unit vectors,we have \\

$$
\begin{array}{cccccccccc}
V_{A:B}^3(\rho)=\{(r_0^1,r_1^1,r_0^2,r_1^2) \in CP^1 \times CP^1:\\
(a_1 r_0^1 r_0^2 +a_7 r_1^1 r_1^2)(a_3 r_0^1 r_1^2 +a_5 r_1^1 r_0^2)(a_4 r_0^1 r_1^2 +a_6  r_1^1 r_0^2)(a_2 r_0^1 r_0^2 +a_8 r_1^1 r_1^2)=0\}
\end{array}
(7)
$$

From Theorem 3 we know that $\rho$ is entangled for the cut A:B:CD, A:C:BD and A:D:BC for generic parameters, since (for example) $a_1r_0^1r_0^2+a_7r_1^1r_1^2$ cannot be factorized to 2 linear forms for generic $a_1$ and $a_7$.  This provides another proof the mixed state is entangled if the 4 parties are isolated.\\

Let $\lambda_1=-a_1/a_7,\lambda_2=-a_3/a_5, \lambda_3=-a_4/a_6, \lambda_4=-a_2/a_8$ and consider the family of the mixed states $\{\rho_{\lambda_{1,2,3,4}}\}$, we want to prove the following statement.\\

{\bf Theorem 4.} {\em The generic memebers in this continuous family of mixed states are inequivalent under the local operations on $H=H_{A}^2 \otimes H_{B}^2 \otimes H_{C}^2 \otimes H_{D}^2$.}\\

{\bf Proof.} From the above computation, $V_{A:B}^3(\rho_{\lambda_{1,2,3,4}})$ is the union of the following 4 algbraic varieties in $CP^1 \times CP^1$.\\

$$
\begin{array}{ccccccccccc}
V_1=\{(r_0^1,r_1^1,r_0^2,r_1^2) \in CP^1 \times CP^1:r_0^1 r_0^2 - \lambda_1 r_1^1 r_1^2=0\}\\
V_2=\{(r_0^1,r_1^1,r_0^2,r_1^2) \in CP^1 \times CP^1:r_0^1 r_1^2 - \lambda_2 r_1^1 r_0^2=0\}\\
V_3=\{(r_0^1,r_1^1,r_0^2,r_1^2) \in CP^1 \times CP^1:r_0^1 r_1^2 - \lambda_3 r_1^1 r_0^2=0\}\\
V_4=\{(r_0^1,r_1^1,r_0^2,r_1^2) \in CP^1 \times CP^1:r_0^1 r_0^2 - \lambda_4 r_1^1 r_1^2=0\}
\end{array}
(8)
$$

From Theorem 2, if $\rho_{\lambda_{1,2,3,4}}$ and $\rho_{\lambda'_{1,2,3,4}}$ are equivalent by a local operation, there must exist 2 fractional linear transformations $T_1, T_2$ of $CP^1$ such that $T=T_1 \times T_2$ (acting on  $CP^1 \times CP^1$) transforms the 4 varieties $V_1,V_2,V_3,V_4$  of $\rho_{\lambda_{1,2,3,4}}$ to the  4 varieties $V'_1,V'_2,V'_3,V'_4$  of $\rho_{\lambda'_{1,2,3,4}}$,ie., $T(V_i)=V'_j$.\\

Introduce the inhomogeneous coordinates $x_1=r_0^1/r_1^1,x_2=r_0^2/h_1^2$. Let $T_1(x_1)=(ax_1+b)/(cx_1+d)$.  Suppose $T(V_i)=V'_i, i=1,2,3,4$. Then we have $ab \lambda_1= cd \lambda'_1 \lambda'_2$ and $ab \lambda_4=cd \lambda'_3 \lambda'_4$. Hence $\lambda_1 \lambda'_3 \lambda'_4 =\lambda'_1 \lambda'_2 \lambda_4$. This means that there are some algebraic relations of parameters if the $T$ exists. Similarly we can get the same conclusion for the other possibilities $T(V_i)=V'_j$. This implies that there are some algebraic relations of parameters $\lambda_{1,2,3,4}$ and $\lambda'_{1,2,3,4}$ if $\rho_{\lambda_{1,2,3,4}}$ and $\rho_{\lambda'_{1,2,3,4}}$ are  equivalent  by a local operation. Hence our conclusion follows immediately.\\

In [22] Nielsen gave a beautiful necessary and sufficient condition for the pure state $|\psi>$ can be transformed to the pure state $|\phi>$ in bipartite quantum systems by local operations and classical communications (LOCC) based on the majorization between the eigenvalue vectors of the partial traces of $|\psi>$ and $|\phi>$. In [10] an example was given, from which we know that Nielsen's criterion cannot be generalized to multipartite case, {\bf 3EPR} and {\bf 2GHZ} are understood as pure states in a $4 \times 4 \times 4$ quantum system, they have the same eigenvalue vectors when traced over any subsystem. However it is proved that they are LOCC-incomparable in [10]\\

In the following example 2, a continuous family $\{\phi\}_{\eta_1,\eta_2,\eta_3}$ of pure states in tripartite quantum system $H_{A_1}^3 \otimes H_{A_2}^3 \otimes H_{A_3}^3$ is given, the eigenvalue vectors of $tr_{A_i}(|\phi_{\eta_1,\eta_2,\eta_3}><|\phi_{\eta_1,\eta_2,\eta_3}>), tr_{A_iA_j}(|\phi_{\eta_1,\eta_2,\eta_3}><\phi_{\eta_1,\eta_2,\eta_3}|)$ are independent of parameters $\eta_1,\eta_2,\eta_3$. However the ``generic'' pure states in this family are entangled and  LOCC-incomparable. This gives stronger evidence that it is  hopeless to characterize the entanglement properties of pure states in multipartite quantum systems by only using the eigenvalue spetra of their partial traces.\\

{\bf Example 2} Let $H=H_{A_1}^3 \otimes H_{A_2}^3 \otimes H_{A_3}^3$ be a tripartite quantum system and the following 3 unit vectors are in $H_{A_1}^3 \otimes H_{A_2}^3$.\\

$$
\begin{array}{cccccccc}
|v_1>=\frac{1}{\sqrt{3}}(e^{i\eta_1}|11>+|22>+|33>)\\
|v_2>=\frac{1}{\sqrt{3}}(e^{i\eta_2}|12>+|23>+|31>)\\
|v_3>=\frac{1}{\sqrt{3}}(e^{i\eta_3}|13>+|21>+|32>)
\end{array}
(9)
$$

,where $\eta_1,\eta_2,\eta_3$ are $3$ real parameters. Let $|\phi_{\eta_1,\eta_2,\eta_3}>=\frac{1}{\sqrt{3}}(|v_1> \otimes |1>+|v_2>\otimes |2>+|v_3> \otimes |3>)$. This is a continuous family of pure states in $H$ parameterized by three real parameters. It is clear that $tr_{A_3}=\frac{1}{3}(|v_1><v_1|+|v_2><v_2|+|v_3><v_3|)$ is a rank 3 mixed state in $H_{A_1}^3 \otimes H_{A_2}^3$. Set $g(\eta_1,\eta_2,\eta_3)=\frac{e^{i\eta_1}+e^{i\eta_2}+e^{i\eta_3}}{e^{i(\eta_1+\eta_2+\eta_3)/3}}$, $|\phi_{\eta_1,\eta_2,\eta_3}>$ and $|\phi_{\eta'_1,\eta'_2,\eta'_3}>$ are not equivalent under local unitary transformations if $k(g(\eta_1,\eta_2,\eta_3)) \neq k(g(\eta'_1,\eta'_2,\eta'_3))$, where $k(x)=\frac{x^3(x^3+216)^3}{(-x^3+27)^3}$ is the moduli function of elliptic curves,   since their corresponding traces over $A_3$ are not equivalent under local unitary transformations of $H_{A_1}^3 \otimes H_{A_2}^3$ from Theorem 5 of [18]. Hence the ``generic'' members of this family of pure states in tripartite quantum system $H$ are enatngled and  LOCC-incomparable from Theorem 1 in [10]. \\

On the other hand it is easy to calculate that all nonzero eigenvalues of $tr_{A_3}(|\phi_{\eta_1,\eta_2,\eta_3}><\phi_{\eta_1,\eta_2,\eta_3}|), tr_{A_1A_3}(|\phi_{\eta_1,\eta_2,\eta_3}><\phi_{\eta_1,\eta_2,\eta_3}|),tr_{A_2A_3}(|\phi_{\eta_1,\eta_2,\eta_3}><\phi_{\eta_1,\eta_2,\eta_3}|)$ are $\frac{1}{3}$. Thus all nonzero eigenvalues of $tr_{A_i}(|\phi_{\eta_1,\eta_2,\eta_3}><|\phi_{\eta_1,\eta_2,\eta_3}>), tr_{A_iA_j}(|\phi_{\eta_1,\eta_2,\eta_3}><\phi_{\eta_1,\eta_2,\eta_3}|)$ are constant $\frac{1}{3}$. Thus the ``generic'' members of this family of pure states have the same eigenvalue spectra but are entangled and LOCC-incomparable.\\ 

For any pure state in a bipartite quantum system $H=H_A^m \otimes H_B^n$ , it can be written as a linear combination of at most $min\{m,n\}$ 2-way orthogonal separable pure states ([10]) from Schmidt decomposition. For multipartite pure states, there is no direct generaliztion of Schmidt decomposition, and those multipartite pure states with a m-way orthogonal decompositions can be distilled to cat states (see [10]). From the results in [11], it is known that we need at least 5 terms of ``orthogonal''  separable pure states to write a generic pure state in $H_A^2 \otimes H_B^2 \otimes H_C^2$ as a linear combination of them. This phenomenon is a remarkable difference between bipartite pure state entanglement and multipartite pure state entanglement. In the following statement it is showed  what happens for generic pure states in $H_A^{n^2} \otimes H_B^{n^2} \otimes H_C^{n^2}$.\\

{\bf Theorem 5.}{\em For a generic pure state $|\psi>=\Sigma_{i=1}^{n^2} \lambda_i |\psi_i^{12}> \otimes |\psi_i^3>$ in a tripartite quantum system $H_A^{n^2} \otimes H_B^{n^2} \otimes H_C^{n^2}$, where $|\psi_1^3>,...,|\psi_{n^2}^3>$ are mutually orthogonal unit vectors in $H_C^{n^2}$ and $|\psi_1^{12}>,...,|\psi_{n^2}^{12}>$ are pure states in $H_A^{n^2} \otimes H_B^{n^2}$, then there exists one of $|\psi_1^{12}>,...,|\psi_{n^2}^{12}>$ with Schmidt rank at least $n$.}\\

{\bf Proof.} It is clear that $tr_C(|\psi><\psi|)=\Sigma_{i=1}^{n^2} |\lambda_i|^2 |\psi_i^{12}><\psi_i^{12}|$ is a generic rank $n^2$ mixed state in $H_A^{n^2} \otimes H_B^{n^2}$. From Theorem 4 in [18], at least one of $|\psi_1^{12}>,...,|\psi_{n^2}^{12}>$ has Schmidt rank (as a pure state in $H_A^{n^2} \otimes H_B^{n^2}$) at least $n$. Thus our conclusion is proved.\\

In conclusion we revealed that there is a large part of multipartite quantum entanglement, which is independent of eigenvalues and only depending on the eigenvectors of the multipartite mixed states. We introduced algebraic set invariants for measuring this part of multipartite quantum entanglement,  which actually are  the {\em degenerating locus} of the Hermitian bilinear forms arising from the measurement of the mixed states by multipartite seaprable pure states. . Based on these algebraic set invariants, a new {\em eigenvalue-free} separability criterion has been  proved. Examples to show why entanglement of tripartite pure states cannot be characterized by the spectra of their partial traces from the point of view of algebraic set invariants and separability criterion have been constructed. Generalized Smolin mixed states were introduced and studied from our invariants and separability criterion, served as examples of continuous many distinct bound entanglement in 4 qubits. We also have proved a conclusion showing the decomposing  a generic tripartite pure state as ``orthogonal'' separable pure states is quite different to the Schmidt decomposition of bipartite pure states.\\

The author acknowledges the support from NNSF China, Information Science Division, grant 69972049.\\

e-mail: chenhao1964cn@yahoo.com.cn\\

\begin{center}
REFERENCES
\end{center}

1.A.Einstein, B.Podolsky and N.Rosen, Phys. Rev. 47,777(1935)\\

2.E.Schrodinger, Proc.Camb.Philos.Soc.,31,555(1935)\\

3.C.H.Bennett, G.Brassard, C.Crepeau, R.Jozsa, A.Peres and W.K.Wootters, Phys.Rev.Lett 70, 1895 (1993)\\

4.C.H.Bennett, G.Brassard, S.Popescu, B.Schumacher, J.Smolin and W.K.Wootters, Phys. Rev.Lett. 76, 722(1996)\\

5.R.Jozsa, in The Geometric Universe, edited by S.Huggett, L.Mason, K.P.Tod, S.T.Tsou, and N.M.J.Woodhouse (Oxford Univ. Press, 1997)\\

6.C.H. Bennett and P.W.Shor, Quantum Information Theory, IEEE Trans. Inform. Theory, vol.44(1998),Sep.\\

7.M.Horodecki, P.Horodecki and R.Horodecki, in Quantum Information--Basic concepts and experiments, edited by G.Adler and M.Wiener (Springer Berlin, 2000)\\

8.D.M.Greenberger, M.Horne and A.Zeilinger, Bell's Theorem, Quantum Theory, and Conceptions of the Universe, edited by M.Kafatos (Kluwer Dordrecht)\\

9.The Physics of Quantum Information, edited by D.Bouwmeester,A.Ekert and A.Zeilinger (Springer, Heidelberg, 2000)\\

10.C.H.Bennett,S.Popescu,D.Rohrlich,J.A.Smolin and A.Thapliyal, Phys. Rev. A 63, 012307, 2000\\

11.A.Acin,A.Andrianov,L.Costa,E.Jane,J.I.Latorre and R.Tarrach, Phys. Rev. Lett. 85, 1560,2001\\

12.N.Linden, S.Popescu and A.Sudbery, Phys.Rev.Lett. 83,243(1999)\\

13.J.Eiskert and H.J.Briegel, Phys. Rev. A 64, 022306, 2001\\

14.A.Peres, Phys. Rev. Lett. 77, 1413 (1996)\\

15.M.Horodecki, P.Horodecki and R.Horodecki, quant-ph/0006071\\

16.C.H.Bennett, D.P.DiVincenzo, T.Mor, P.W. Shor, J.A.Smolin and T.M. Terhal, Phys.Rev. Lett. 82,5385 (1999)\\

17.A.Acin,D.Bruss, M.Lewenstein and A.Sanpera, Phys. Rev. Lett. 87, 040401(2001)\\

18.Hao Chen, Quantum entanglement without eigenvalue spectra, preprint July,2001, quant-ph/0108093\\

19.J.Harris, Algebraic geometry, A first Course, Gradute Texts in Mathematics, 133, Springer-Verlag, 1992,especially its Lecture 9  ``Determinantal Varieties''\\

20.J.A.Smolin,Phys.Rev. A 63, 032306(2001)\\

21.M.Lewenstein,D.Bruss,J.I.Cirac,B.Krus,J.Samsonowitz,A.Sanpera and R.Tarrach,J.Mod.Optics,47,2481 (2000),quant-ph/0006064\\

22.M.A.Nielsen, Phys. Rev. Lett. 83,436(1999)\\

23.Hao Chen, Quantum entanglement and geometry of determinantal varieties, quant-ph/0110103\\

\end{document}